%% file: main.tex
\documentclass{article}

\usepackage{arxiv}

\usepackage[utf8]{inputenc} 
\usepackage[T1]{fontenc}    
\usepackage{hyperref}       
\usepackage{url}            
\usepackage{booktabs}       
\usepackage{amsfonts}       
\usepackage{amsmath}
\usepackage{blkarray}
\usepackage{nicefrac}       
\usepackage{microtype}      
\usepackage{lipsum}	
\usepackage{graphicx}
\usepackage{natbib}
\usepackage{doi}
\usepackage{todonotes}
\usepackage[title]{appendix}
\usepackage[export]{adjustbox}
\usepackage{array}

\title{Discriminatory or Samaritan -- which AI is needed for humanity? An Evolutionary Game Theory Analysis of Hybrid Human-AI populations}

\author{
 	Tim Booker* \\
	Complexity Science Hub Vienna, Austria \\
	\texttt{\href{mailto:booker@csh.ac.at}
    {booker@csh.ac.at}} \\
    \And
 	Manuel Miranda* \\
	IFISC \\
	UIB-CSIC \\
	\texttt{\href{mailto:jeslop@ifisc.uib-csic.es}{mmiranda@ifisc.uib-csic.es}} \\
    \And
 	Jesús A. Moreno López* \\
	IFISC \\
	UIB-CSIC \\
	\texttt{\href{mailto:jeslop@ifisc.uib-csic.es}{jeslop@ifisc.uib-csic.es}} \\
    \And
    José María Ramos Fernández* \\
	Universidad de La Laguna\\
	\texttt{\href{mailto:}{alu0101100883@ull.edu.es}} \\
	\And
	Max Reddel*\\
	Department of Emerging Technology Governance\\
	International Center for Future Generations, Belgium\\
	\texttt{\href{mailto:max@reddel.ai}{max@reddel.ai}} \\
	\And
  	Valeria Widler* \\
	Modeling and Simulation of Complex Processes \\
	Zuse Institute Berlin \\
	\texttt{\href{mailto:}{widler@zib.de}} \\
	\And
   	Filippo Zimmaro* \\
	Department of Mathematics, University of Bologna\\
	Department of Computer Science, University of Pisa \\
	\texttt{\href{mailto:}{filippo.zimmaro@phd.unipi.it}} \\
    \And
   	Alberto Antonioni \\
	Department of Mathematics \\
	University Carlos III de Madrid | UC3M \\
	\texttt{\href{mailto:aantonio@math.uc3m.es}{aantonio@math.uc3m.es}} \\
    \And
    The Anh Han**\\
    	School of Computing, Engineering and Digital Technologies,\\
    	Teesside University, UK\\
    	\texttt{\href{mailto:T.Han@tees.ac.uk}{T.Han@tees.ac.uk}}
     \And
     * These authors contributed equally.
     ** Corresponding author.
}

\renewcommand{\shorttitle}{Human-AI-Hybrid Cooperation}

\hypersetup{
pdftitle={PDF Title},
pdfsubject={bla},
pdfauthor={Max Reddel},
pdfkeywords={First keyword, Second keyword, More},
}

\begin{document}
\maketitle

\input{sections/abstract}
\newpage

\input{sections/intro}

\input{sections/models_and_methods}
\input{sections/results}
\input{sections/discussion}

\input{sections/conclusion}

\clearpage

\input{sections/appendix}

\end{document}

%% file: sections/abstract.tex
\begin{abstract}
As artificial intelligence (AI) systems are increasingly embedded in our lives, their presence leads to interactions that shape our behaviour, decision-making, and social interactions. 
Existing theoretical research has primarily focused on human-to-human interactions, overlooking the unique dynamics triggered by the presence of AI.
In this paper, resorting to methods from evolutionary game theory, we study how different forms of AI influence the evolution of cooperation in a human population playing the one-shot Prisoner's Dilemma game in both well-mixed and structured populations.
We found that Samaritan AI agents that help everyone unconditionally, including defectors, can promote higher levels of cooperation in humans than Discriminatory AI that only help those considered worthy/cooperative, especially in slow-moving societies where change is viewed with caution or resistance (small intensities of selection). Intuitively, in fast-moving societies (high intensities of selection), Discriminatory AIs promote higher levels of cooperation than Samaritan AIs.
\end{abstract}

%% file: sections/intro.tex
\section{Introduction}

Societies are dynamic networks of interconnected individuals, consistently engaging in a multitude of interactions~\citep{stella2018bots}. Quite frequently, these individuals, or agents, aspire to enhance their personal well-being. However, this aspiration often interlocks with the need for collective welfare, thereby creating a classic social dilemma. This dilemma poses a conflict between personal self-interests and the shared good of society. Addressing such cooperative problems represents a recurring, intricate challenge at every level of human society~\citep{dasgupta2023investigating,perc2017statistical}. These challenges span a wide array of issues, from the negotiation of the national budgets to managing communal spaces in shared residences, dealing with homelessness, and confronting global predicaments such as climate change and financial crises.

In the digital world, human agents are increasingly accompanied by autonomous entities with various intelligence levels, thus creating a more complex social system. These Artificial Intelligence (AI) agents have unique capabilities, such as disseminating information and recommending content, which can potentiate both cooperative and selfish behaviors in human agents~\citep{fernandez2022delegation}.
One of the concerning instances of AI deployment is in large-scale social manipulation, often referred to as "social hacking". As an illustration, a study by~\citep{stella2018bots} on the 2017 Catalan independence referendum offered insightful findings. This research explored the "structural and emotional roles played by social bots" in catalyzing the polarization between Independentists and Constitutionalists on Twitter. The investigation employed extensive social data and uncovered that bots targeted influential members of both factions, bombarding Independentists with content designed to incite outrage. This strategy effectively amplified the conflict between the two groups.
Similarly, during the 2016 US presidential election, it's estimated that over 400,000 bots were actively participating in political discussions on Twitter~\citep{keijzer2021strength}. These bots had a considerable role in shaping public discourse, thereby exemplifying the broad and impactful influence AI agents can exert in our digital society.

On the other hand, intelligent agents pose the unique potential to promote social intelligence, i.e. the capacity of collectives to effectively cooperate in solving shared problems~\citep{dafoe2020open,dasgupta2023investigating}.
As the share of human-AI interactions grows and the capabilities of intelligent agents evolve, the presence of AI in human decision-making is increasingly impactful in shaping the trade-offs of social dilemmas and, therefore, our capacity to achieve common goals.
The emergent consequences of these hybrid-AI interaction dynamics remain largely obscure and under-researched. Consequently, it is pivotal to deepen our understanding of the impact of human-AI dynamics on strategic decision-making and collective action problems.
As human-AI interactions become more common and the capabilities of intelligent agents evolve, AI could shape the outcomes of social dilemmas, and therefore, our capacity to achieve common goals, via its influence on human decision-making. 
The emergent consequences of these hybrid-AI interaction dynamics remain largely obscure and under-researched. Consequently, deepening our understanding of the impact of human-AI dynamics on strategic decision-making and collective action problems is pivotal.

In this paper, we build an evolutionary game theory (EGT)~\citep{key:Hofbauer1998} model to study the impact of different forms of AI agents in a population of humans playing the one-shot Prisoner's Dilemma game. 
It allows us to study how AI impacts the balance between cooperation and selfish behaviour, and how the presence of AI could facilitate prosocial behaviour among humans. 
Research into the emergence and maintenance of cooperative behaviour has been a key concern for disciplines as varied as game theory, sociology, multi-agent systems, physics and psychology~\citep{tuyls2007evolutionary,perc2017statistical,key:Sigmund_selfishnes} for decades. 
Illustrating the conflict between individual and collective welfare decision-making, it is sometimes described as “the perfect metaphor for the study of the evolution and emergence of cooperation among large populations”~\citep{xianyu2012prisoner}. 
Thus, it provides a suitable frame to study the impact of AI on pro-social behaviour. 

Furthermore, we examine the impacts of different strategies employed by AI systems. Specifically, we analyse two approaches: Samaritan AI, which consistently exhibits positive behaviour by choosing to cooperate at each interaction, and Discriminatory AI, which rewards pro-social actions and penalises anti-social ones. We assume that these artificial agents possess the capability to recognize the intentions of human agents they interact with.
In this context, Discriminatory AI agents collaborate with those who cooperate and act against those who defect. The primary research question revolves around determining the most effective strategy, either Samaritan or Discriminatory, in promoting cooperation among the human population. Furthermore, we investigate how their effectiveness is influenced by various factors, such as game parameters, the underlying network of interactions, and the governing dynamic rule. To provide a complete analysis, we also consider an anti-social strategy called Malicious AI, which consistently defects.


For a comprehensive understanding, we adopt complementary approaches in our analysis, including replicator dynamics for infinite populations~\citep{key:Sigmund_selfishnes} and stochastic simulations for finite populations~\citep{nowak:2006bo}, both for well-mixed population settings. 
Moreover, to explore the impact of network structures, we also use agent-based modelling simulations in networks.  

The research paper is organized as follows. First, we present the model and methods used in our study. This includes evolutionary dynamics within well-mixed populations, the implementation of agent-based simulations, and an overview of the adopted network topologies.
Following this section, we provide analytical and numerical results for finite and infinite populations.
The subsequent section is dedicated to the discussion of the results, offering a comprehensive interpretation of the findings and their potential implications.
The paper concludes by summarizing the study's findings and delineating potential paths for future research.
Finally, we provide appendix sections including additional results.

%% file: sections/models_and_methods.tex
\section{Model and Methods}
In the following, we describe the hybrid human-AI models and methods used to analyse the models.  

\subsection{Model}

We consider a population interacting on three networks: a complete graph (equivalent to a well-mixed population), a square lattice, and a Barabási–Albert scale-free network. Interactions between agents are a one-shot Prisoner's Dilemma (PD) game, with the following pay-off matrix:
\[
\begin{blockarray}{ccc}
    & C & D\\
    \begin{block}{c(cc)}
      C & R,R & S,T \\
      D & T,S & P,P \\
    \end{block}
  \end{blockarray}.
\]
If both interacting players follow the same strategy, they receive the same reward: $R$ if it is mutual cooperation and $P$ if it is mutual defection. If the agents play different strategies, the cooperator gets the sucker's pay-off $S$, and the defector gets the temptation to defect $T$. The pay-off matrix corresponds to the preferences associated with the PD when the parameters satisfy the ordering $T>R>P>S$. The PD can be reduced to its special case, the donation game, by considering $T = b$, $R = b-c$, $P = 0$ and $S = -c$, where $b$ and $c$ are the benefit and cost of cooperation, respectively.

In this work, we study AI agents' influence on a population of humans who play the one-shot Prisoner's Dilemma. To do so, we examine three different cases of AI behaviour. First, AI behaviour is fixed as either cooperative or defective. We refer to this as \textbf{AI = C} or \textbf{AI = D}.   Second, AIs can predict human behaviour, and account for this in its decision in the PD interaction. For example, AI might take into account the human's reputation. We refer to this behaviour as Intention Recognition, or \textbf{AI = IR}, as in \citep{han2015synergy, dafoe2020open,key:hanetalAdaptiveBeh}. In our analysis, we assume IR has perfect intention recognition for simplicity. The general case of imperfect intention recognition (non-zero failure probability) is described in the Appendix. As a ``prosocial'' AI, it will cooperate if its interaction partner cooperates or has a good reputation and defect otherwise.

We also refer to AIs  (AI = C) and (AI = IR) as \textbf{Samaritan AI} and \textbf{Discriminatory AI}, respectively. 


\begin{table}
\centering
\begingroup
\renewcommand{\arraystretch}{1.25}
\begin{tabular}{lc}
\hline
 Parameter  &  Symbol 
 \\ 
\hline
 Number of human agents & $N$ 
 \\ Number of AI  agents & $M$ 
 \\
  Cost of cooperation & $c$ 
 \\
 Benefit of cooperation & $b$ 
 \\
 Intensity of selection & $\beta$  
 \\
Prisoner's dilemma payoff entries   & R, S, T, P   \\
\\
\end{tabular}
\endgroup
 \caption{Summary of model parameters }
 \label{table:parameters}
\end{table}

\subsection{Methods}

\subsubsection{Evolutionary dynamics in well-mixed populations}

The agents' payoffs represent their \textit{social success} or \textit{fitness}, while the evolutionary dynamics are shaped by social learning \citep{key:Hofbauer1998,key:Sigmund_selfishnes}, according to what the agents of the populations will tend to imitate the most successful ones. In this project, social learning is modelled using the pairwise comparison, by which an agent A with fitness $f_{A}$ adopts an agent's B strategy with a probability $p$ given by the Fermi rule \citep{traulsen2006}
\[
p_{A,B} = (1+e^{-\beta(f_{B}-f_{A})})^{-1},
\]
where $\beta$ represents the "imitation strength" or "intensity of selection" that accounts for how much the agents base their choice to imitate on the difference in the fitness between the opponents and themselves. Thus, the imitation is random if $\beta = 0$, and it becomes increasingly deterministic for large values of $\beta$.

We consider that if a human agent is aware that the role model is an AI agent, it might have a different probability of imitating the strategy  of AI. We model that by distinguishing between  human's and AI's  intensities of selection. They would be denoted as $\beta_H$ and $\beta_{AI}$, respectively, where appropriate (see Appendix A).

Let us consider a well-mixed population of $N$ humans and $M$ AIs. For all the agents, there are only two possible strategies: cooperate C and defect D. In the first scenario, we assume that humans can either cooperate or defect and change between strategies while AI agents have fixed behaviour; for example, they always cooperate. In this context, consider that at a given moment, there are $k$ human agents with strategy $C$ and so $(N-k)$ human agents using strategy $D$. If we denote by $\pi_{x,y}$ the payoff that a strategist $x$ obtains in an interaction with strategist $y$, we can define the payoff of both agents $C$ and $D$ as follows

\begin{equation}
\begin{split}
\Pi_{C}(k) = \frac{(k-1)\pi_{C,C}+(N-k)\pi_{C,D}+M\pi_{C,AI}}{N+M-1},\\
\Pi_{D}(k) = \frac{k\pi_{D,C}+(N-k-1)\pi_{D,D}+M\pi_{D,AI}}{N+M-1}.
\end{split}
\end{equation}
where  the payoffs for interactions between human and AI are given as  follows 
\begin{equation}
     \pi_{H,AI} = \lbrace\begin{array}{cc}          \pi_{H,C} &~if~AI=C\\
          \pi_{H,D} &~if~AI=D\\
          \pi_{H,H} &~if~AI=IR
     \end{array}
\end{equation}

\begin{equation}
     \pi_{AI,H} = \lbrace\begin{array}{cc}          \pi_{C,H} &~if~AI=C\\
          \pi_{D,H} &~if~AI=D\\
          \pi_{H,H} &~if~AI=IR
     \end{array}
\end{equation}

Substituting the rewards with the parameters of the general Prisoner's dilemma, in the well-mixed case we get
\begin{equation}
\Pi_{C}(k) = \frac{(k-1)R+(N-k)S+M(\delta_{_{AI,C}}R + \delta_{_{AI,D}} S + \delta_{_{AI,IR}}R)}{N+M-1}
\end{equation}

\begin{equation}
\Pi_{D}(k) = \frac{kT+(N-k-1)P+M(\delta_{_{AI,C}}T + \delta_{_{AI,D}} P + \delta_{_{AI,IR}}P)}{N+M-1}. 
\end{equation}

On the other hand, the probability of increasing/decreasing by one the number $k$ of human agents using strategy $C$ in each time step is given by

\begin{equation}
\label{eq:Tplusminus}
    \begin{split}
    T^{+}(k) &= \frac{N-k}{k}\frac{k+(1-\delta_{_{AI,IR}})\delta_{_{AI,C}}\cdot M}{N+M}\left[1+e^{-\beta(\Pi_{C}-\Pi_{D})}\right]^{-1},\\
    T^{-}(k) &= \frac{k}{N}\frac{N-k+(1-\delta_{_{AI,IR}})(1-\delta_{_{AI,C}})M}{N+M}\left[1+e^{\beta(\Pi_{C}-\Pi_{D})}\right]^{-1},
\end{split}
\end{equation}

where the first term in both expressions gives the probability of choosing a human defector or cooperator, respectively, the second term gives the probability of selecting an agent, either human or AI, that cooperates or defects, respectively. We have used, to distinguish the payoffs and the transition probabilities related to the three possible AI behaviours, calling $St \;= \;C,D,IR$ a general strategy

\[
\delta_{_{AI,St}} = \left\lbrace\begin{array}{cc}
1&if~AI=St \\
0& \ \text{otherwise}
\end{array}\right.
\]

The fixation probability of a single mutant with a strategy $A$ in a population of $(N-1)$ agents using $B$ is given by \citep{traulsen2006,key:novaknature2004}
\begin{equation} 
\label{eq:fixprob} 
\rho_{B,A} = \left(1 + \sum_{i = 1}^{N-1} \prod_{j = 1}^i \frac{T^-(j)}{T^+(j)}\right)^{-1}.
\end{equation} 

Considering a set  $\{1,...,q\}$ of different strategies, these fixation probabilities determine a transition matrix $M = \{T_{ij}\}_{i,j = 1}^q$, with $T_{ij, j \neq i} = \rho_{ji}/(q-1)$ and  $T_{ii} = 1 - \sum^{q}_{j = 1, j \neq i} T_{ij}$, of a Markov Chain. The normalised eigenvector associated with the eigenvalue 1 of the transposed of $M$ provides the stationary distribution described above \citep{key:imhof2005}, describing the relative time the population spends adopting each of the strategies.








\subsection{Agent-based Simulations}


\subsubsection{Network structures}

Well-mixed populations offer a convenient baseline scenario where a specific heterogeneous interaction structure is absent (all interact with all). By studying structured populations, we go one step beyond and ask whether network properties and structural heterogeneity can foster the evolution of guilt-prone behaviours. To begin with, we study square lattice (SL) populations of size $N = 50 \times 50$ --- a  widely adopted population structure in population dynamics and evolutionary games (for a survey, see \citep{Szabo2007}), wherein each agent can only interact with its four immediate neighbours (aside from edge and corner nodes). With the SL, we introduce a network structure where all nodes can be considered structurally equivalent. 

Next, we explore complex networks in which the network portrays a heterogeneity that mimics the power-law distribution of wealth (and opportunity) characteristic of real-world settings. The Barab{\'a}si and Albert (BA) model \citep{barabasi1999emergence} is one of the most famous models used in the study of such heterogeneous, complex networks. The main features of the BA model are that it follows a \textit{preferential attachment} rule, has a small clustering coefficient, and a typical \textit{power-law degree distribution}. To explain preferential attachment, let us describe the construction of a BA network. Starting from a small set of $m_0$ interconnected nodes, each new node selects and creates a link with $m$ older nodes according to a probability proportional to their degree (number of its edges). The procedure stops when the required network size of $N$ is reached. This will produce a network characterised by a power-law distribution, $p_{k} \sim k^{-\chi}$, where the exponent $\chi$ is its $\textit{degree exponent}$ \citep{Barabasi2016}. There is a high degree correlation between nodes, and the degree distribution is typically skewed with a long tail. There are few hubs in the network that attract an increasing number of new nodes, which attach as the network grows (in a typical \textit{``rich-get-richer''} scenario). The power-law distribution exhibited by BA networks resembles the heterogeneity present in many real-world networks. The average connectivity of the resulting scale-free network is $z = 2m$. For our experiments, we pre-seed ten different scale-free networks of size $N = 1000$, with average connectivity of $z = 4$, to coincide with the number of neighbours in a square lattice.

\subsubsection{Computer Simulations}
Initially, $M$ randomly selected nodes in the network are assigned to be AI. Each of the remaining $N-M$ agents in the network are designated C or D, with equal probability. At each time step, each agent plays the PD with its neighbours. The fitness score for each agent is the sum of the payoffs in these encounters. At the end of each step, an agent $A$ with fitness $f_A$ chooses to copy the strategy of a randomly selected neighbour agent $B$ with score $f_B$, with a probability given by the Fermi function  \citep{Szabo2007}, as given above. Similar to the well-mixed setting above, we set $\beta = 1$ in our simulations. 

We simulate this evolutionary process until a stationary state or a cyclic pattern is reached. 
For a transparent and fair comparison, all simulations are run for $10^5$ steps. We use an asymmetric update approach, i.e. only one individual is eligible for imitation at each time step.
Moreover, for each simulation, the results are averaged over the last $1000$ generations to account for the fluctuations characteristic of these stable states. 
Furthermore, to improve accuracy, for each set of parameter values, the final results are obtained by averaging 30 independent runs (for SF, ten seeded networks with 20 runs each, i.e. 200 runs in total).

\subsection{Replicator dynamics}

The previous approach is useful when the number of agents (humans and AI) is not too big. But difficulties may arise as the size of the system increases, because the range of values that the number of cooperators $k$ may take increases. For that purpose, we will also consider a system with infinite number of agents. For that purpose, instead of using $N$ and $M$ as the number of human agents and AI agents, we will denote by $x$ the fraction of cooperators among humans and by $\alpha = \frac{M}{N+M}$ to the fraction of AI among the total number of agents, which will remain fixed as the size of the population tends to infinity. This approach allows us to analyze the behavior of the system in the context of a continuous population and explore the effects of AI agents on strategy dynamics.

With this considerations, we can benefit from the fact that there exist only two strategies to write the evolution of the fraction of human cooperators:

\begin{equation}
    \dot x = (1-x)\left[x(1-\alpha)p_{D,C}(\beta_{H}) + \alpha p_{D,C}(\beta_{AI})\right]- x(1-x)(1-\alpha)p_{C,D}(\beta_{H}),
\end{equation}

where $p_{C,D}(\beta_H)$ and $p_{D,C}(\beta_H)$ is the probability of a human imitating the strategy of another human with opposite strategy, following the Fermi rule described before with parameter $\beta_H$. 

On the other hand, $p_{D,C}(\beta_{AI})$ is the probability of a human defector to imitate the behavior of an AI. As we will study two AI behaviors, this probability, as well as the average payoff, will change. For instance, if we consider Samaritan AI, this probability will follow the Fermi rule similar to the probability of imitating another human, with a possibly different parameter $\beta_{AI}$. Notice that, if the value $\beta_{AI} \not= \beta_H$, it means that humans can differentiate humans from AI. If we consider a Discriminatory AI, this probability is simply zero. 

%% file: sections/results.tex
\section{Results}

   \subsection{Analytical results for finite populations} 
    Calculating the stationary distribution (for the current model with two strategies $C$ and $D$), through the fixation probabilities $\rho_{D,C}$ and $\rho_{C,D}$, we obtain \citep{nowak:2006bo,han2018cost}
    \[
    \frac{\rho_{D,C}}{\rho_{C,D}+\rho_{D,C}} = \frac{ r } {1+r},  
    \]
    where  $r = \frac{\rho_{D,C}}{\rho_{C,D}}= \prod_{k = 1}^{N-1} \frac{T^+(k)}{T^-(k)}$. In the following, we determine $r$ for  cases, corresponding to the three different behaviours of the AI agents. We consider $\beta_{AI}= \beta_{H}:= \beta$, and note that in all cases  
    \[
    \frac{p_{D,C}(\beta) }{p_{C,D}(\beta)} = e^{\beta \Delta f(k)}, 
    \]
    where $k$ refers to the number of human cooperators and $\Delta f(k) = \Pi_C(k) - \Pi_D(k)$ depends on the AI behaviour. We rewrite the fraction  
    \[
         \frac{\rho_{D,C}}{\rho_{C,D}} = G_{AI}(N,M) e^{\beta F_{AI}(N,M)},
    \]
    where $AI= C$, D, or IR, and we determine the functions $G_{AI}(N,M)$ and $F_{AI}(N,M)$ in the three cases. Afterwards, we determine when cooperation is risk dominant against defection, i.e. when $\frac{\rho_{D,C}}{\rho_{C,D}} >1$, which is equivalent to 
    \[
         H_{AI}(N,M) :=  \beta F_{AI}(N,M) - \log{\bigg(\frac{1}{G_{AI}(N,M)} \bigg)} > 0.
    \]
    We find that 

    \begin{align}
        & F_{C}(N,M) = \frac{N-1}{N+M-1}\left[\Big((P-R)+N(S-P)+M(R-T)\Big)+(R+P-T-S)\frac{N}{2}\right] \\
        & F_D(N,M) = \frac{N-1}{N+M-1}\left[\Big((P-R)+N(S-P) + M(S-P)\Big)+(R+P-T-S)\frac{N}{2}\right] \\
        & F_{IR}(N,M) = \frac{N-1}{N+M-1}\left[\Big((P-R)+N(S-P)+M(R-P)\Big) + (R+P-T-S)\frac{N}{2}\right] \\
        &\\
        & G_C(N,M) = \frac{(N-1+M)!}{(N-1)!M!}\\
        & G_D(N,M) = \frac{(N-1)!(M+1)!}{(N-1+M)!}\\
        & G_{IR}(N,M) = 1        
    \end{align}    

    For the $AI = C$
    \[
    \frac{T^{+}(k)}{T^{-}(k)} = \frac{k+M}{k}\frac{P_{DC}(\beta)}{P_{CD}(\beta)}
    \]
    and so
    \[
    \frac{\rho_{DC}}{\rho_{CD}} = \prod_{k = 1}^{N-1}\frac{k+M}{k}e^{\beta\Delta f(k)} = \left(\prod_{k=1}^{N-1}\frac{k+M}{k}\right)e^{\beta\sum_{k = 1}^{N-1}\Delta f(k)}
    \]
    from which we have 
    \[
    G_C(N,M) = \prod_{k=1}^{N-1}\frac{k+M}{k}  = \frac{(N-1+M)!}{(N-1)!M!}
    \]
    In this case the payoffs read
    \[
    \begin{split}
    \Pi_{C}(k) = \frac{(k-1)R+(N-k)S+MR}{N+M-1}, \\
    \Pi_{D}(k) = \frac{kT + (N-k-1)P + MT}{N+M-1}.
    \end{split}
    \]
    Thus,
    \[
    \Delta f(k) = \frac{1}{N+M-1}\left[k(R-T)+(N-k)(S-P)+(P-R)+M(R-T)\right]
    \]
    and
    \[
    F_{C}(N,M) = \sum_{k=1}^{N-1} \Delta f(k)  = \frac{1}{N+M-1}\left[(N-1)\Big((P-R)+N(S-P)+M(R-T)\Big)+(R+P-T-S)\frac{N(N-1)}{2}\right].
    \]
    For the $AI = D$
    \[
    \frac{T^{+}(k)}{T^{-}(k)} = \frac{N-k}{N-k+M}\frac{P_{DC}(\beta)}{P_{CD}(\beta)}
    \]
    and so
    \[
    \frac{\rho_{DC}}{\rho_{CD}} = \prod_{k = 1}^{N-1}\frac{N-k}{N-k+M}e^{\beta\Delta f(k)} = \left(\prod_{k=1}^{N-1}\frac{N-k}{N-k+M}\right)e^{\beta\sum_{k = 1}^{N-1}\Delta f(k)}
    \]
    from which we have 
    \[
    G_D(N,M) = \prod_{k=1}^{N-1} \frac{N-k}{N-k+M}=  \frac{(N-1)!(M+1)!}{(N-1+M)!}
    \]
    in this case the payoffs read
    \[
    \begin{split}
        \Pi_{C}(k) = \frac{(k-1)R+(N-k)S+MS}{N+M-1},\\
        \Pi_{D}(k) = \frac{kT+(N-k-1)P+MP}{N+M-1}.
    \end{split}
    \]
    thus
    \[
        \Delta f(k) = \frac{1}{N+M-1}\left[k(R-T)+(N-k)(S-P)+ (P-R) + M(S-P)\right]
    \]
    and
    \[
        F_D(N,M) := \sum_{k=1}^{N-1} \Delta f(k) = \frac{1}{N+M-1}\left[(N-1)\Big((P-R)+N(S-P) + M(S-P)\Big)+(R+P-T-S)\frac{N(N-1)}{2}\right]. 
    \]

    For $AI = IR$
    \[
    \frac{T^{+}(k)}{T^{-}(k)} = \frac{P_{DC}(\beta)}{P_{CD}(\beta)}
    \]
    and so
    \[
    \frac{\rho_{DC}}{\rho_{CD}} = e^{\beta\sum_{k=1}^{N-1}\Delta f(k)}.
    \]
   Thus, in this case, we can simplify to 
    \[
    G_{IR}(N,M) = 1.
    \]
    In this case the payoffs read
    \[
    \begin{split}
        \Pi_{C}(k) = \frac{(k-1)R+(N-k)S+MR}{N+M-1}, \
        \Pi_{D}(k) = \frac{kT + (N-k-1)P + MP}{N+M-1}.
    \end{split}
    \]
    Thus,
    \[
    \Delta f(k) = \frac{1}{N+M-1}\left[k(R-T) + (N-k)(S-P) + (P-R) + M(R-P)\right]
    \]
    and
    \[
    F_{IR}(N,M) = \sum_{k = 1}^{N-1}\Delta f(k) = \frac{1}{N+M-1}\left[(N-1)\Big((P-R)+N(S-P)+M(R-P)\Big) + (R+P-T-S)\frac{N(N-1)}{2}\right].
    \]

\begin{figure}
    \centering
    \includegraphics[width=1.2\linewidth]{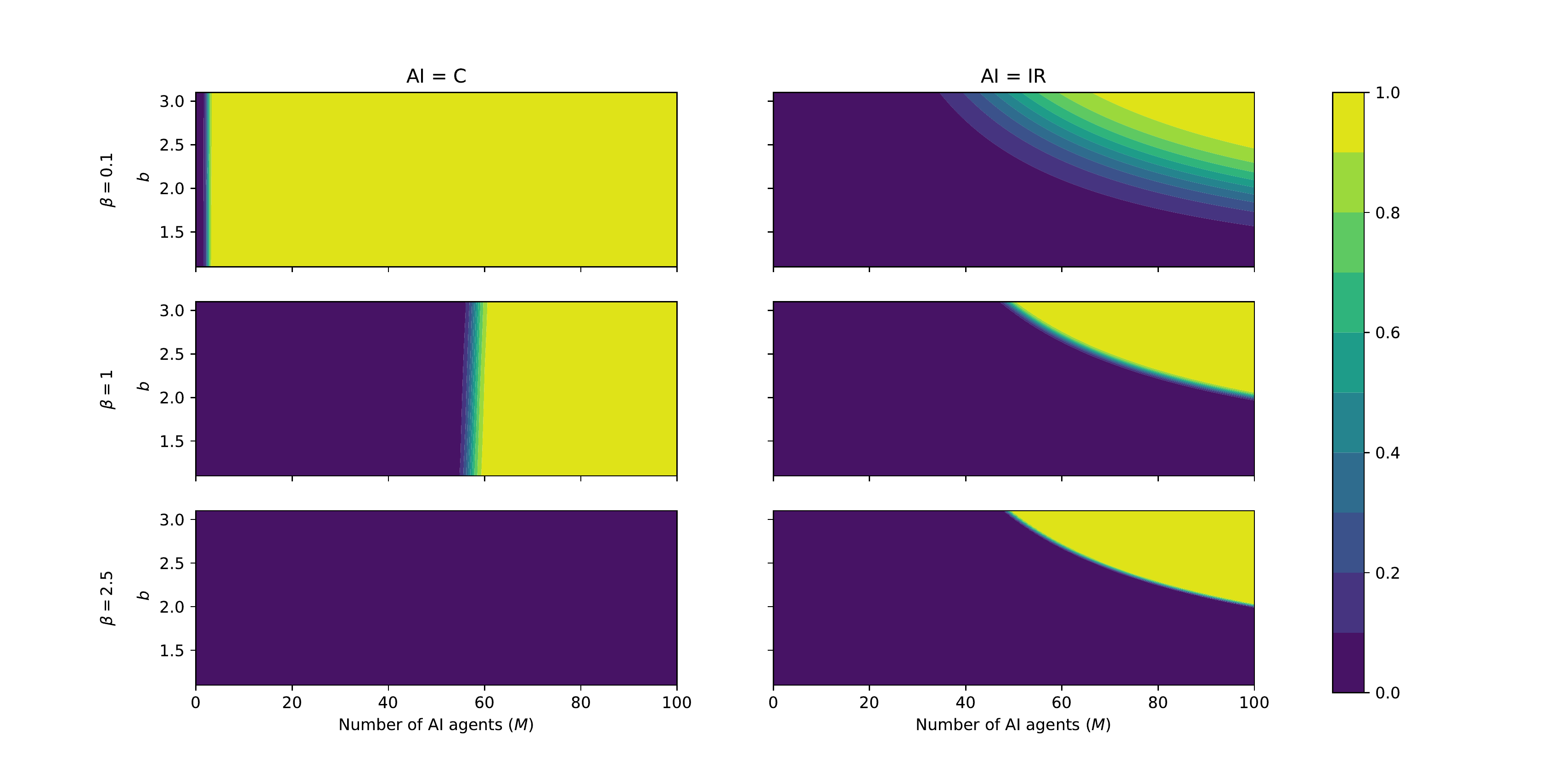}
    \caption{\textbf{Samaritan AI promotes higher levels of cooperation than discriminatory AI for weak intensities of selection, and vice versa for stronger intensities of selection.} We show the frequency of cooperation obtained for varying  number of AIs, $M$,  and benefit of cooperation, $b$, for Samaritan AI (left column, AI = C) and Discriminatory AI (AI = IR). Each row is associated to a different value of $\beta$. Other parameters: population size $N = 100$.}
    \label{anal_results_fin_pop}
\end{figure}

\subsection{Numerical results for finite populations}
In Figure \ref{anal_results_fin_pop} we calculate the frequency of cooperation (see Methods) for   varying  the number of AIs and  and benefit of cooperation, for both Samaritan AI  and Discriminatory AI, for different regimes of intensities of selection. 
The results show that, when $\beta$ is small (first row, $\beta = 0.1$, also see Appendix for lower $\beta$),  Samaritan AI can leads to significantly higher levels of cooperation than  Discriminatory AI, even with fewer AI agents. When $\beta$ is high (third row), Discriminatory AI is better at promoting cooperation. 

The results imply that, in a slow moving society where behaviour change or imitation is viewed with caution and resistance,  Samaritan AI agents that help everyone regardless of their behaviour, is more conductive for enhancing cooperation in humans that they interact with. The presence of the Samaritan AIs enables more cooperation through increasing the chance that defectors meet a cooperator as a role model (the second term of $T^+$, see Equation \ref{eq:Tplusminus}), which out-weights the payoff benefit  provided by Discriminatory AIs (for cooperators vs defectors). 
On the other hand, in a fast-moving society (characterized by fast-faced dynamics, i.e. high $\beta$), the payoff benefit enabled by Discriminatory AIs for cooperators is significantly magnified (in the Fermi functions, see Equation \ref{eq:Tplusminus}) and thus out-weights the benefit of Samaritan AIs as role models.

   \subsection{Analytical results for infinite populations}
    We work in the regime $N\rightarrow\infty$ and $\frac{M}{N+M}\rightarrow\alpha$ ($\alpha$ is the fraction of $AI$ agents in the system), in the well-mixed topology. We write and analyze the evolution equations for the cases of $AI=C$ and $AI=IR$. \\
    \\
    Considering a time-unit corresponding to $N$ steps of the dynamics, in the case $AI=C$ we have that the fraction of cooperators among humans, i.e., $x= \frac{k}{N}$, evolves as
    \begin{align}
        \dot x &= (1-x)\left[x(1-\alpha)p_{D,C}(\beta_{H}) + \alpha p_{D,C}(\beta_{AI})\right]- x(1-x)(1-\alpha)p_{C,D}(\beta_{H})\\
        &= (1-x) \big[ x(1-\alpha)( p_{D,C}(\beta_{H}) - p_{C,D}(\beta_{H})  ) + \alpha  p_{D,C}(\beta_{AI})  \big]\\
        &= (1-x) \big[ x(1-\alpha)\tanh{\bigg(\frac{\beta_H \Delta f(x)}{2}\bigg)} +  \frac{\alpha}{1+ e^{-\beta_{AI}\Delta f(x)}} \label{eq:EvolutionReplicator} \big]
    \end{align}
    where
    \[
    \Delta f(x) = \Pi_C(x) - \Pi_D(x) = (R-T)[(1-\alpha)x+\alpha] + (S-P)(1-\alpha)(1-x) 
    \]
    is the difference in payoffs, always negative in this case.\\
    Clearly, the evolution \eqref{eq:EvolutionReplicator} has at least one fixed point at $x^*=1$, corresponding to full cooperation. Another fixed point appears whenever $ h(x) := x(1-\alpha)\tanh{\bigg(\frac{\beta_H \Delta f(x)}{2}\bigg)} + \alpha  p_{D,C}(\beta_{AI}) = 0$ has a solution $x^*\in [0,1)$. We can derive a sufficient condition for $h(x)=0$ noticing that it always holds that $h(0)>0$, and by Bolzano's theorem if $h(1)<0$ at least one solution in the range exists.

    We have that $h(1)<0$ when 
    \[
    \alpha < \frac{1}{1+\frac{1}{(1+e^{\beta_{AI}(T-R)})\tanh{(\frac{\beta_{H}}{2}(T-R))}}} := \alpha_c
    \] 
    In the case $\beta_{AI} = \beta_{H}= \beta$, the expression simplifies
    \[
    \alpha_{c} = \frac{e^{-\frac{ \beta}{2} (T-R)}}{e^{\frac{\beta}{2}(T-R)}+e^{-\frac{\beta}{2}(T-R)}}
    \]

\begin{figure}
        \centering
        \includegraphics[width=0.49\linewidth]{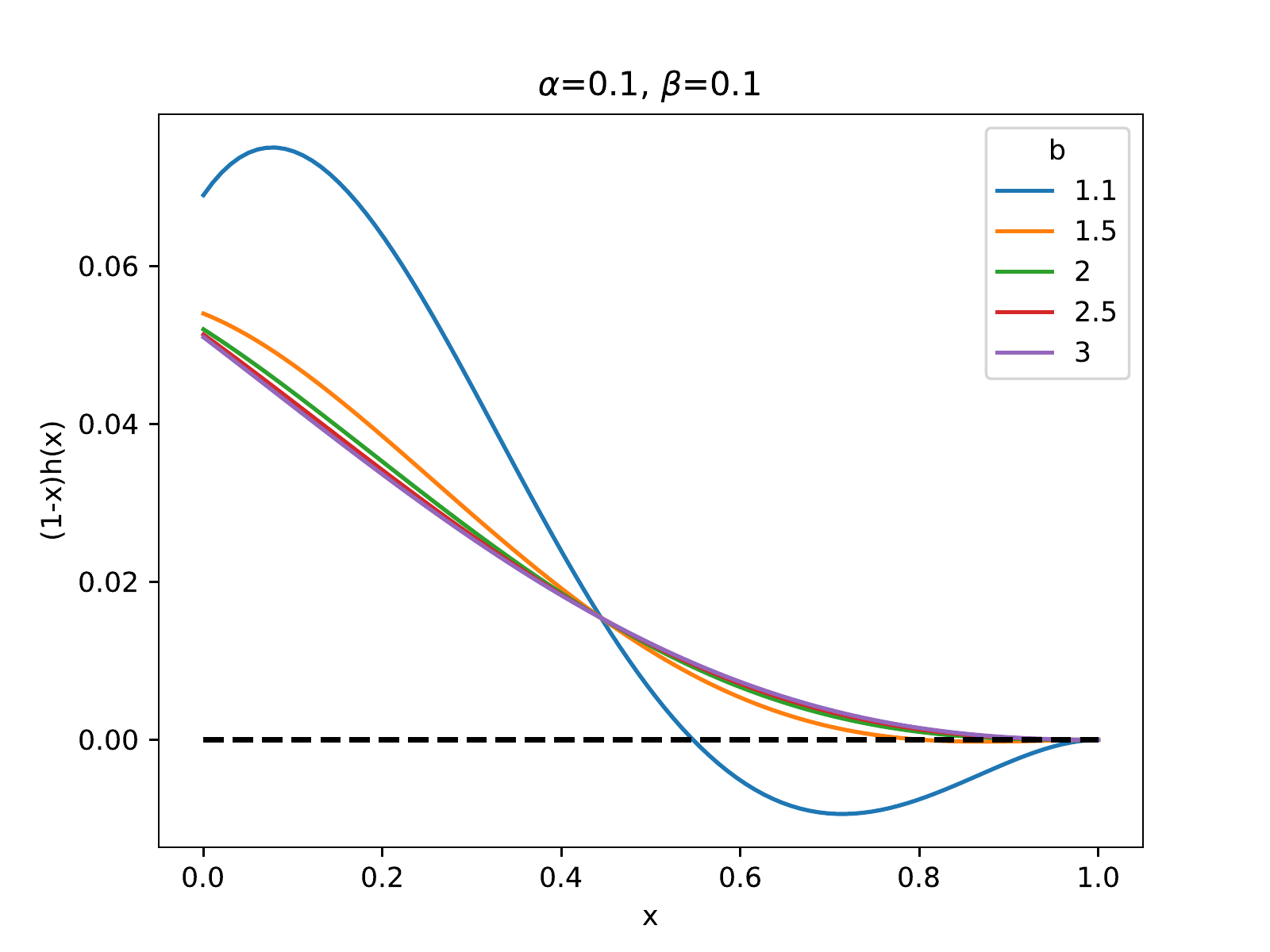}
        \includegraphics[width=0.49\linewidth]{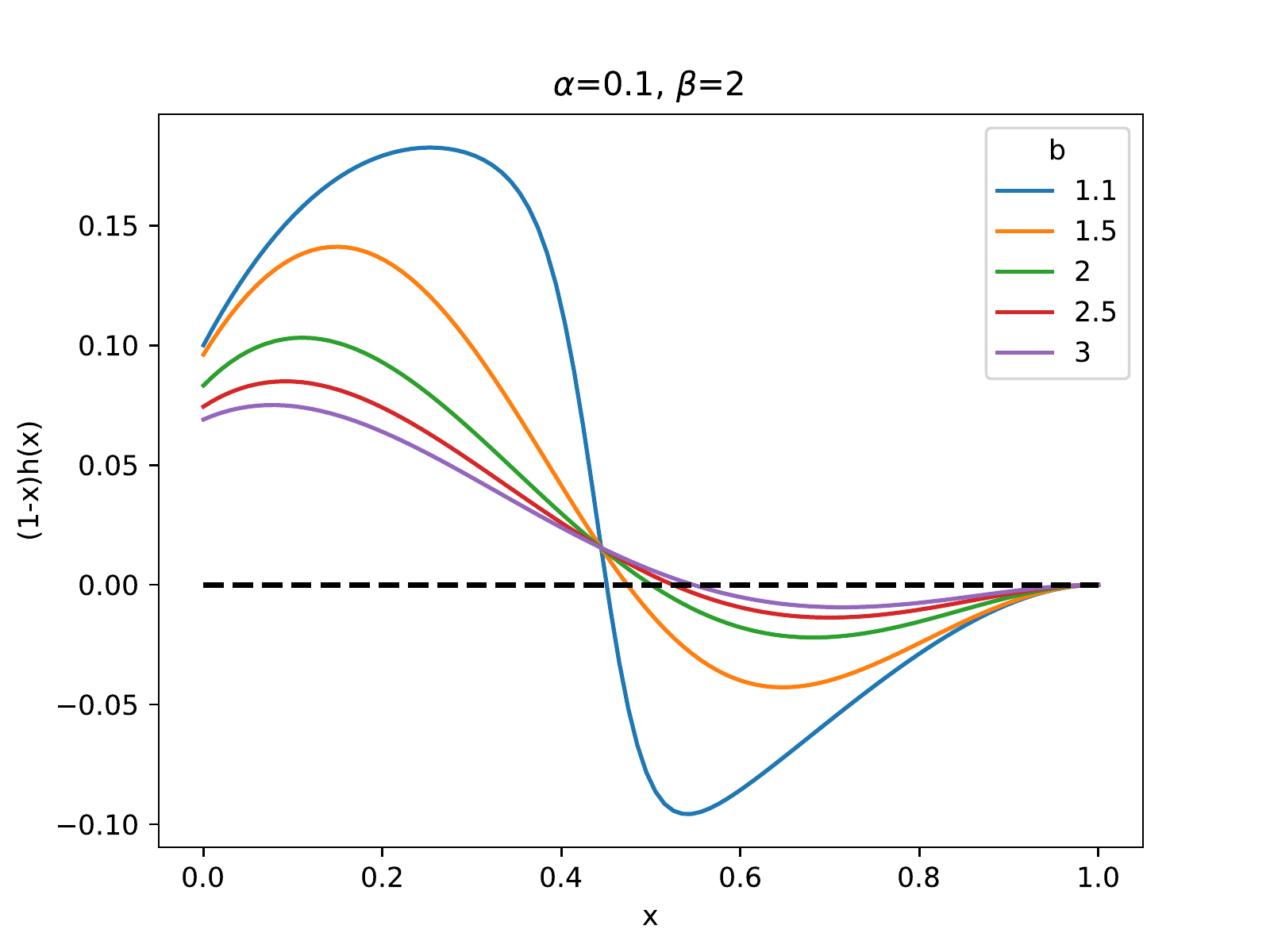}
        \caption{Function $(1-x)h(x)$ for different values of the benefit $b$. Left figure shows the results for low values of $\beta$ while the right one shows the function for high values, both of them for $\alpha=0.1$ fixed.}
        \label{fig:hfunctionb}
    \end{figure}

\begin{figure}
        \centering
        \includegraphics[width=0.49\linewidth]{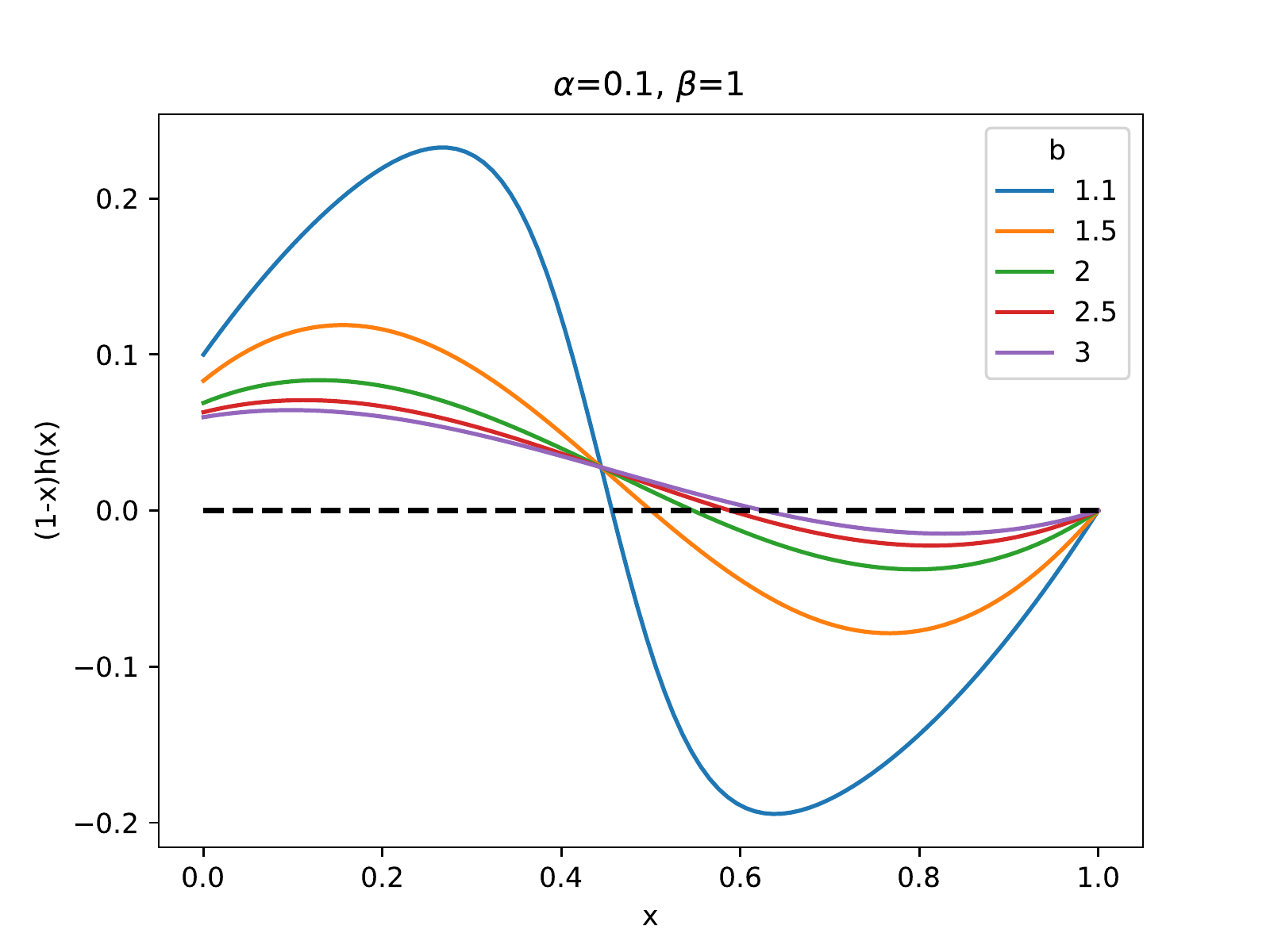}
        \includegraphics[width=0.49\linewidth]{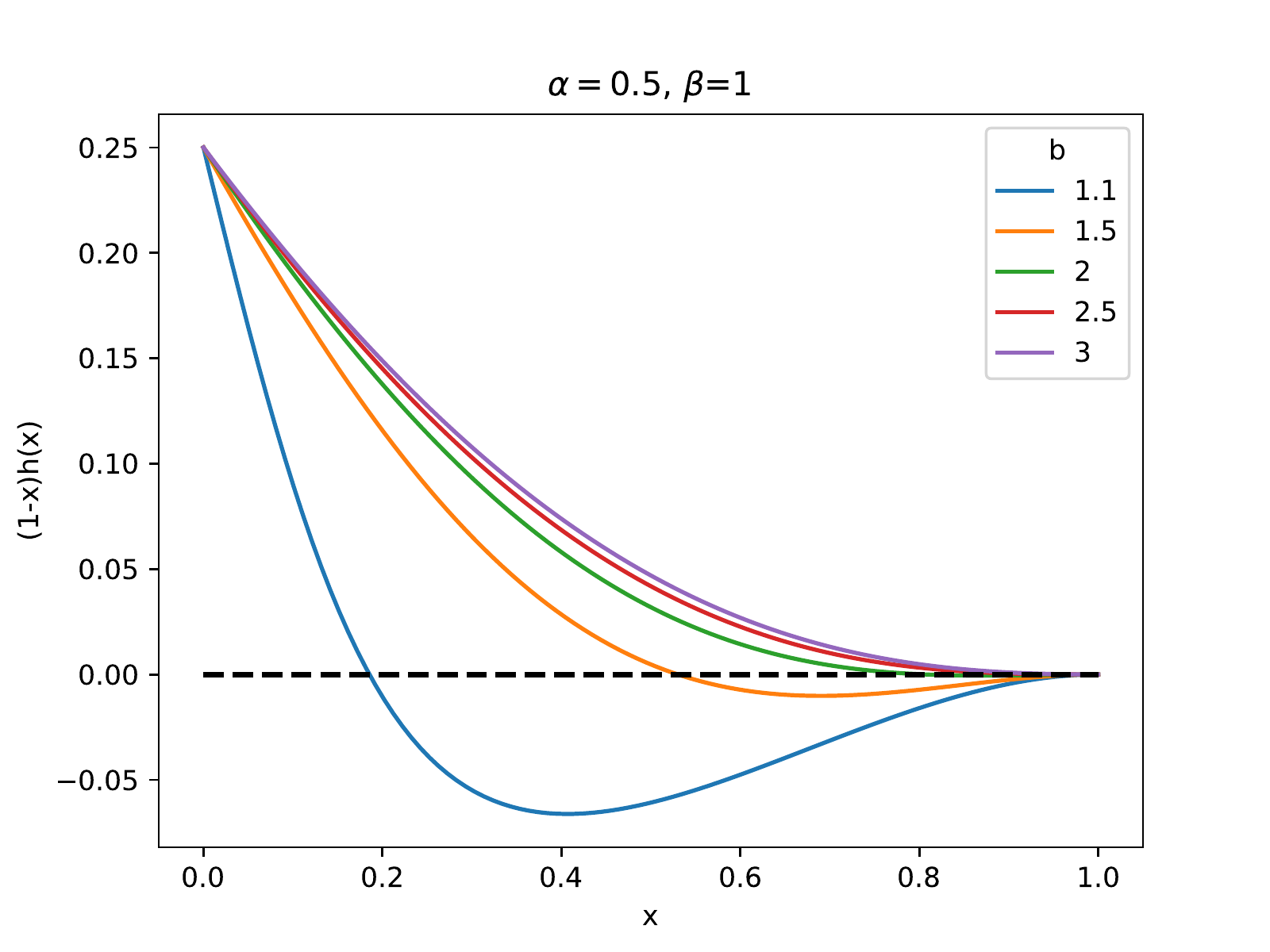}
        \caption{Function $(1-x)h(x)$ for different values of the benefit $b$ and two different fractions $\alpha$ of AIs in the system.}
        \label{fig:hfunctionab}
    \end{figure}

    Considering a donation game, we plot the function $(1-x)h(x)$ in Figures \ref{fig:hfunctionb} and \ref{fig:hfunctionab} for different values of $\alpha$, $\beta=\beta_H=\beta_{AI}$ and $b$, where $b$ is the benefit in the donation game for a fixed cost $c=1$.

    In Figure \ref{fig:hfunctionb} we can observe that the existence of non-full cooperation fixed points depends highly on the parameter $\beta$. In slowly-changing societies with low values of $\beta$, humans always tend to cooperate for lower benefits $b$. On the other hand, fast societies with higher $\beta$ need to be incentivized in order to cooperate. Nevertheless, high enough incentive leads to all humans cooperating, even in this case.
    
    In Figure \ref{fig:hfunctionab} we observe the behavior of the function for fixed $\beta=1$. We can see how the fraction of AIs changes the fixed point for all benefits, and we observe that, for low fraction of AIs, there is an stable point $0<x<1$ even for high benefits, while when we increase $\alpha$ to $0.5$ we quickly observe that $0$ and $1$ are the unique fixed points.

    Numerical simulations suggest that there exists and $\Tilde{\alpha}$ for which, if $\alpha > \Tilde{\alpha}$, then the only fixed point of Equation \eqref{eq:EvolutionReplicator} is $x^*=1$. This is, when the fraction of Samaritan AI is high enough, the only stable strategy for humans is to cooperate all.

    
    \vskip 1 cm
    
    In the case $AI=IR$, the evolution equation reads
    \begin{align}
        \dot x &= (1-x)\left[x(1-\alpha)p_{D,C}(\beta_{H})\right]- x(1-x)(1-\alpha)p_{C,D}(\beta_{H})\\
        &= (1-x)x (1-\alpha) \tanh{\bigg(\frac{\beta_H \Delta f(x)}{2}\bigg)}.  
    \end{align}
    Notice that in this case there cannot be transitions for a human when interacting with an AI, as each AI always copies his current strategy: the AI presence enters only in the payoff difference, which reads
    \[
    \Delta f(x) = \Pi_C(x) - \Pi_D(x) = (R-T)(1-\alpha)x + (S-P)(1-\alpha)(1-x) + \alpha(R-P)
    \]
    and whose sign depends on the parameters and the fraction of cooperators. In this case, together with the two fixed points $x^*=0,1$, there can be another one if $h(x)= 0$ has a solution $x^* \in (0,1)$.

    
    
    

   \subsection{Agent-based simulation results for square lattice}
    
\begin{figure}[!ht]
    \includegraphics[width=\linewidth]{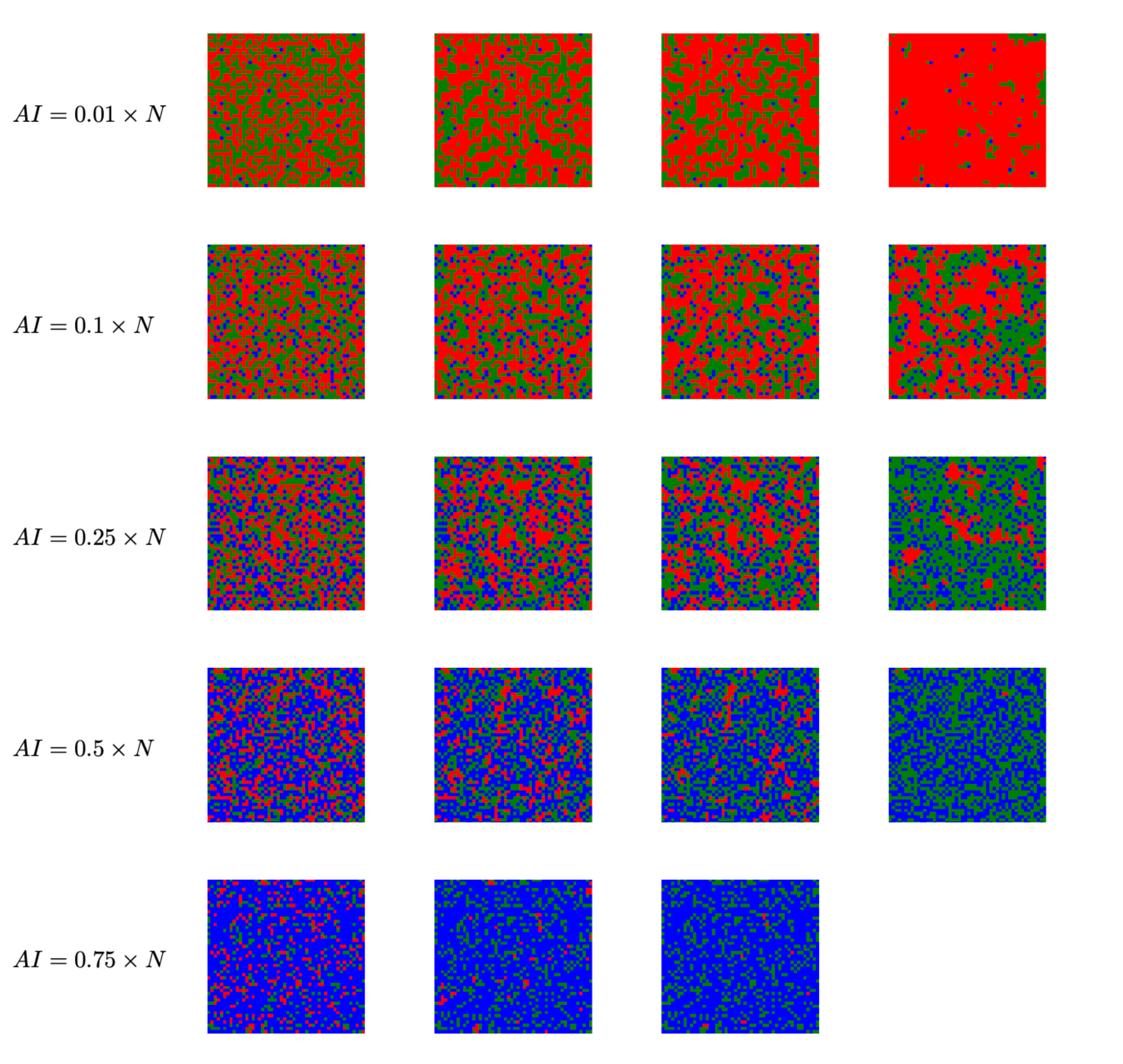}
    \caption{\label{SL-beta=0.1} \textbf{ Snapshots for the evolution of cooperator and defector in the square lattice network} at generation $0$, $5000$, $10,000$, $100,000$, respectfully, for various proportions of AI agents. Green, red, and blue squares denote cooperators, defectors, and AI agents, respectively. Parameters:  
    $\beta=0.1, \ b = 2, \ c =1.$}
    \end{figure}

\begin{figure}[!ht]

\includegraphics[width=\linewidth]{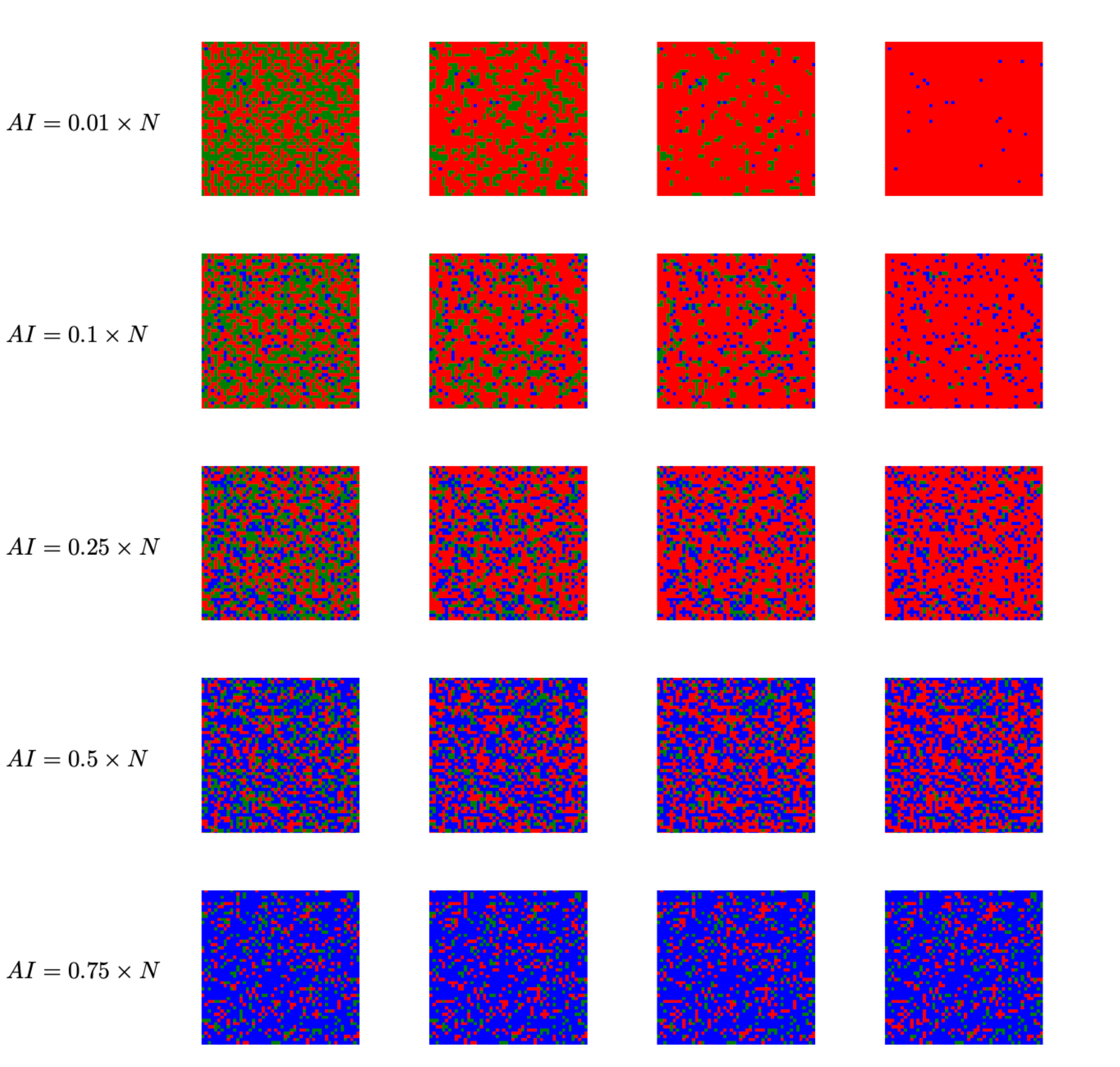}
   
        \caption{\label{SL-beta=1}Snapshots of the square lattice network with $\beta=1$,  with the same values for other parameters as in Figure 4.}
        \end{figure}

\begin{figure}[!ht]
     
        \includegraphics[width=\linewidth]{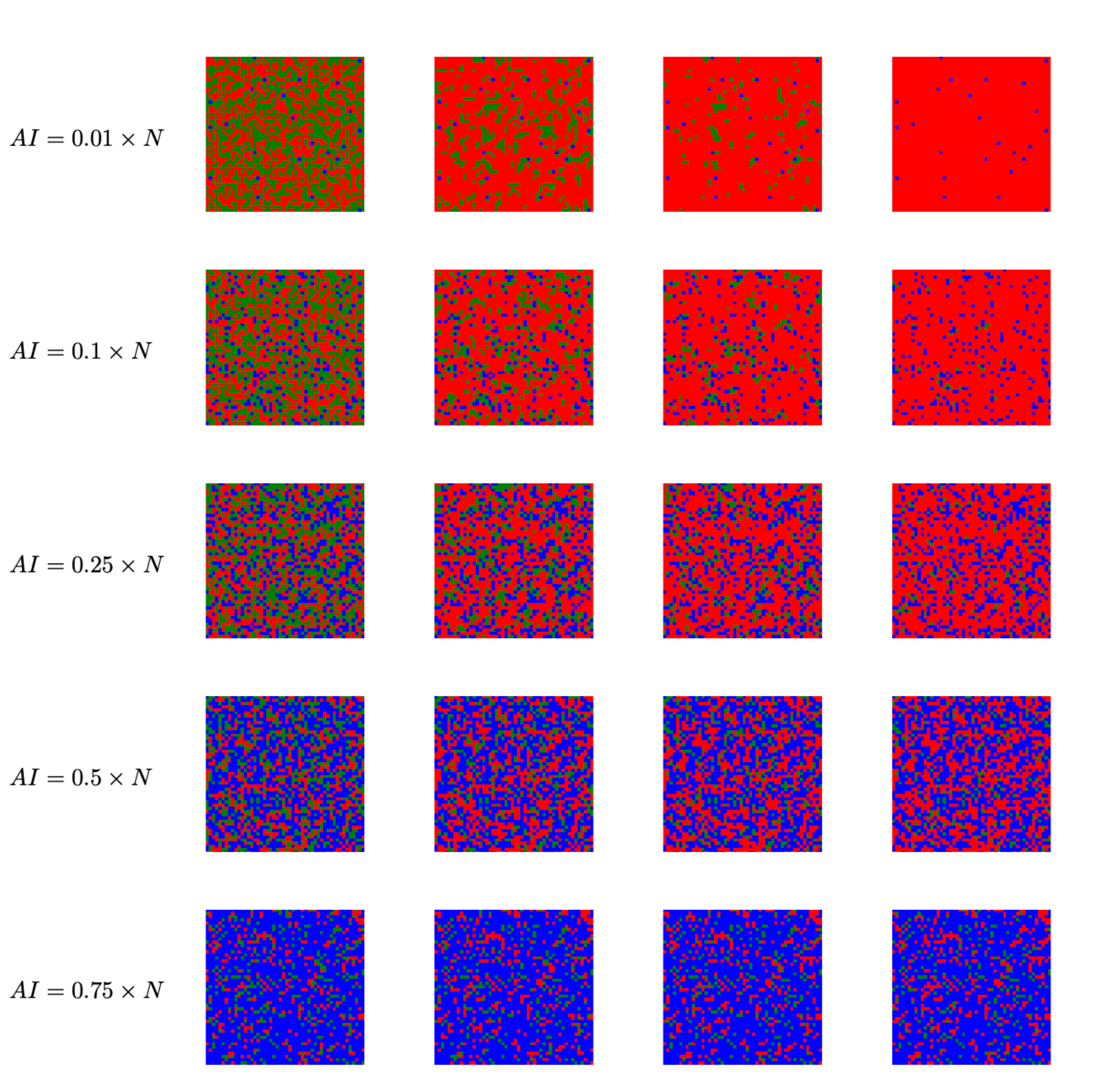}
        \caption{\label{SL-beta=5}Snapshots of the square lattice network with $\beta=5$, with the same values for other parameters as in Figure 4.}
        \end{figure}

    In Tables \ref{SL-beta=0.1}, \ref{SL-beta=1} and \ref{SL-beta=5} we show the evolution over time of the strategies in the population, for $\beta = 0.1$, $1.0$
and $5$, respectively. For each case, we consider different fractions of AI in the population. 
   Similarly to the observations in the well-mixed case (Section 3.2), the results show that Samaritan AI promote high levels of cooperation for low intensities of selection, which also increase when the fraction of AIs is higher.  
   
     \subsubsection{Scale-free Networks}
\begin{figure}[!ht]
    \includegraphics[width=\linewidth]{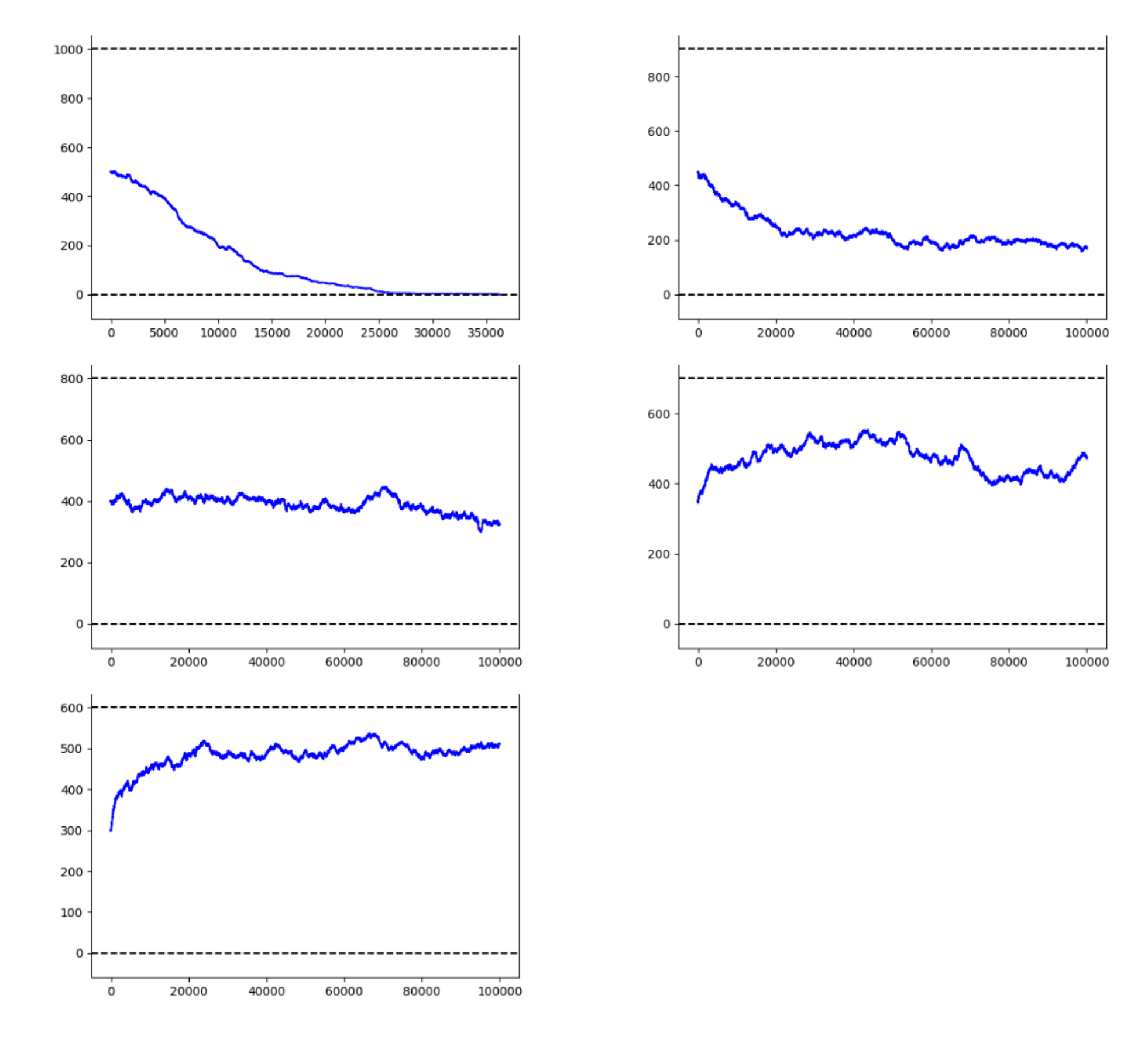}
        \caption{\label{BA-beta=0.1}Time evolution of cooperator count for BA network for various AI proportions (0, 0.1, 0.2, 0.3, 0.4 (left to right, top to bottom)). Model parameters: $b=2, c = 1, \ \beta=0.1.$}
        \end{figure}

\begin{figure}[!ht]
        \includegraphics[width=\linewidth]{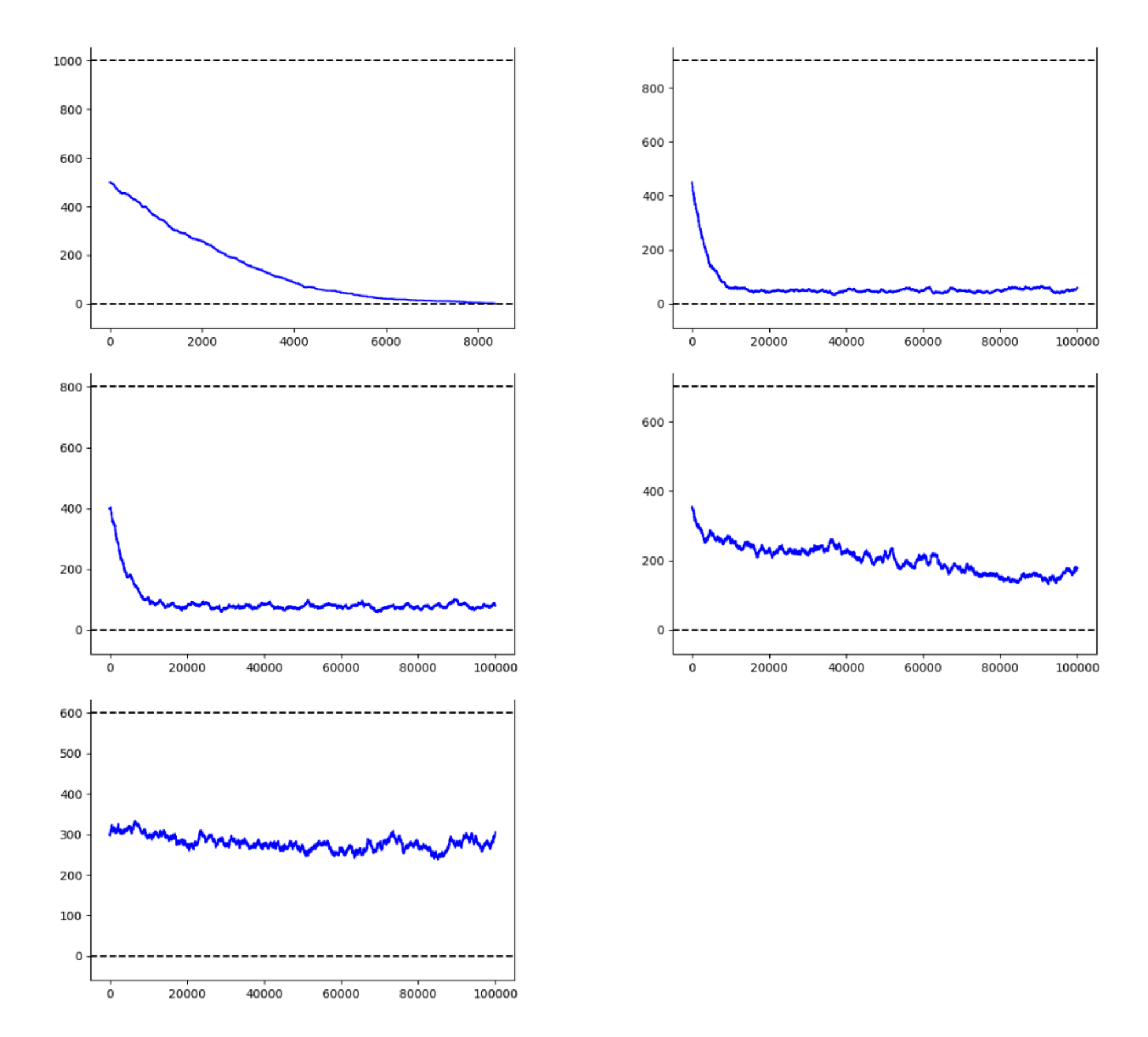}
        \caption{\label{BA-beta=1}Time evolution of cooperator count for BA network for various AI proportions (0, 0.1, 0.2, 0.3, 0.4 (left to right, top to bottom)). Model parameters: $b=2, \beta=1.$}
        \end{figure}

\begin{figure}[!ht]
\includegraphics[width=\linewidth]{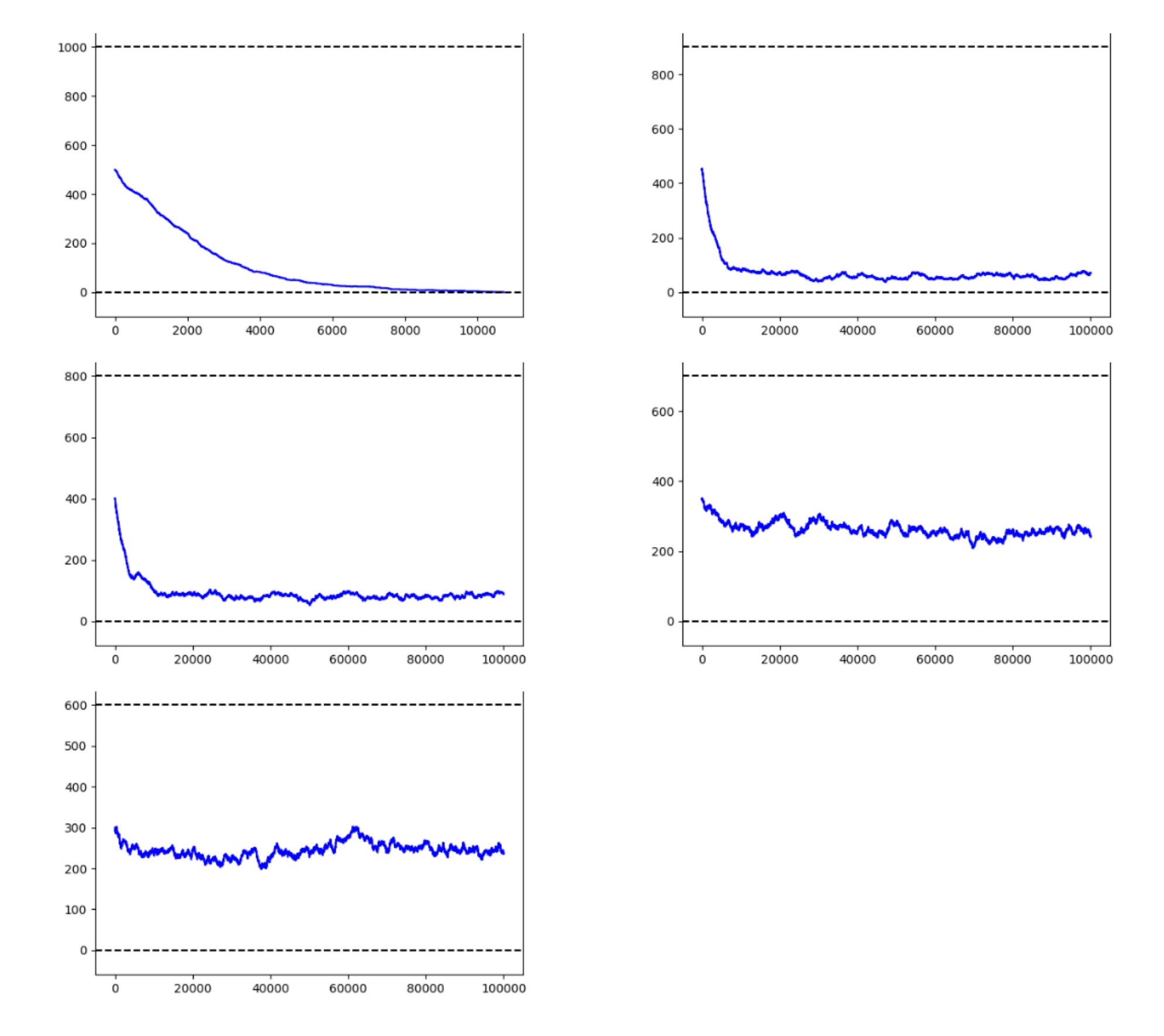}
        \caption{\label{BA-beta=5}Time evolution of cooperator count for BA network for various AI proportions (0, 0.1, 0.2, 0.3, 0.4 (left to right, top to bottom)). Model parameters: $b=2, \beta=5.$}
        \end{figure}

    In Figures \ref{BA-beta=0.1}, \ref{BA-beta=1} and \ref{BA-beta=5}, we show preliminary results of the time evolution of the number of cooperators in the network for increasing proportions of AI agents. We use the same model parameters as in the SL, $\beta=0.1, 1$ and $5$, and $b=2$, $c=1$. Likewise, agent roles are initialised randomly and uniformly across the network. The BA network is generated with an average degree of $4$, coinciding with the SL. 
    
    The results align with that of the SL network. Samaritan AIs promote high levels of cooperation when there is a low level of selection intensity and only slightly prevent complete defection for high levels of selection intensity. The results being similar across two network structures is consistent with network topology not influencing the final state of the model. However, we stress that the exact distribution of AI agents, particularly with regard to their occupancy of hub nodes, could drastically influence the evolution of the network. Experiments with agent distributions biased towards hub nodes remain to be conducted. 
    
%% file: sections/discussion.tex
\newpage
\section{Discussion}
According to research on cooperative AI by \citep{dafoe2020open}, the cooperative equilibrium can be achieved via mechanisms that create a conditional commitment, communication or the ability to predict other players' behavior. For example, commitments can be implemented by repeated play, the costs of one’s reputation, a contract or a physical constraint. Another method is to ‘change the game’ which can be achieved by the creation of institutions. These institutions are effective in facilitating collective cooperation, if they promote cooperative understanding, communication or commitments. What all of these methods share, is the mutual identification of cooperating players, called positive associativity \citep{pepper2002mechanism}.

In the case of a one-shot game with no memory and reputation mechanisms, positive associativity is not possible. The cooperative regime can be achieved nonetheless by way of rewarding altruists, the optional PD, where players can also opt-out, punishing defectors, as well as the introduction of zealot players \citep{sharma2023small}.  

The punishment of defectors can be achieved by way of third party punishment or altruistic punishment. According to \citep{fehr2004third}, third party punishment is directly related to the enforcement of norms that regulate human behavior. If a first party, or defector, harms a second party, or cheated player, we can introduce a third party that is unaffected by the violation and who can impose a punishment on the norm violating player. The action, however, is costly for the third party and does not incur a reward. For a rational agent, it would be illogical to engage in punishing the first party at a personal cost. In other words, the third party punishment presents a dilemma of social cooperation \citep{fehr2004third}. 
In the case of our model, however, the third party is played by an AI, that does not bear the cost of punishment and effectively solves the dilemma. 
Contrary to basic tenets of evolutionary game theory, zealous players are a type of agent that does not imitate other players' strategy but remains committed regardless of payoffs \citep{nakajima2015evolutionary}. The mechanism that shifts the equilibrium of defectors towards cooperation with a certain number of zealous cooperators, is the increased probability cooperative individuals to encounter other cooperators. However, according to this effect can only be observed in well-mixed populations \citep{sharma2023small}. 

To investigate the effectiveness of autonomous intelligent agents for pro-social behavior, we implemented zealot players as Samaritan AI and third-party punishment with Discriminatory AI respectively on the facilitation of pro-social regimes in evolutionary one-shot Prisoner's Dilemma in hybrid-AI systems, on a well-mixed population, square lattice and Barabási-Albert network structure.

The results imply that, in a slow moving society where behaviour change or imitation is viewed with caution and resistance,  Samaritan AI agents that help everyone regardless of their behaviour, is more conductive for enhancing cooperation in humans that they interact with. The presence of the Samaritan AIs enables more cooperation through increasing the chance that defectors meet a cooperator as a role model (the second term of $T^+$, see Equation \ref{eq:Tplusminus}), which out-weights the payoff benefit  provided by Discriminatory AIs (for cooperators vs defectors). 
On the other hand, in a fast-moving society (characterized by fast-faced dynamics, i.e. high $\beta$), the payoff benefit enabled by Discriminatory AIs for cooperators is significantly magnified (in the Fermi functions, see Equation \ref{eq:Tplusminus}) and thus out-weights the benefit of Samaritan AIs as role models.


%% file: sections/conclusion.tex
\section{Conclusion}
Our in-depth analysis has yielded noteworthy insights into the substantial influence exerted by the intensity of selection concerning the type of AI agents that can best promote the emergence of improved human cooperation. By focusing on the intensity of selection, we aim to capture the degree to which AI agent characteristics impact the evolution of cooperation among human agents in a given population. Our findings suggest that the type of AI agent, be it Samaritan, Discriminatory, or Malicious, can profoundly shape cooperation dynamics, especially when the intensity of selection is high. When the intensity of selection is strong, traits that confer even a slight advantage are likely to be increasingly prevalent over time. In the context of our study, this implies that the selection of certain types of AI agents can either promote or inhibit the development of cooperative behavior among human agents, depending on the specific behaviors these AI agents exhibit. Thus, our research underscores the potential for carefully designed AI agents to encourage enhanced cooperation within human societies, providing a novel perspective on managing social dilemmas.

While our model offers useful insights into the role of AI agents in promoting human cooperation, it does have some limitations that could be addressed in future work. First, our model assumes that AI agents have a perfect understanding of human intentions. Future models could incorporate more realistic assumptions about the cognitive capabilities of AI.
Secondly, our model focuses on a static network of interactions. In reality, social networks evolve over time, with relationships forming and dissolving. Future iterations of this model could benefit from incorporating the dynamism of real-world social networks.
Finally, our model explores only three types of AI agents. The universe of AI strategies is vast, and different types of AI could have different impacts on human cooperation. Future research could expand the types of AI agents studied, perhaps investigating more nuanced AI strategies.
By addressing these issues, we can enhance the accuracy of our model and provide even more robust insights into the role of AI in promoting human cooperation.

To confirm the predictions made by our model, it's crucial to execute large-scale behavioural experiments across various countries and cultures. The Samaritan and Discriminatory AI strategies we've discussed can be analyzed in diverse social and cultural contexts to truly understand their effectiveness in promoting cooperation. Given that people exhibit different behavioural types~\citep{kurzban2005experiments, balietti2021optimal, szekely2021evidence} influenced by their cultural background, personal beliefs, experiences and psychological traits, these variations need to be considered. AI agents should be designed to recognize these individual traits and adjust their behaviours accordingly. For instance, an AI agent might employ a more Samaritan strategy with individuals who respond well to cooperation, while taking a more Discriminatory approach with those who exhibit less cooperative behaviours. This personalized approach could enhance the ability of AI agents to promote cooperation across a broad spectrum of human behaviours, leading to more nuanced and effective AI-human interactions.



\subsection*{Acknowledgments}
This work is the output of the Complexity72h workshop, held at IFISC in Palma, Spain, 26-30 June 2023. \hyperlink{https://www.complexity72h.com}{https://www.complexity72h.com}

%% file: sections/appendix.tex
\clearpage

\begin{appendices}

\input{sections/formulas}

\counterwithin{figure}{section}
\renewcommand\thefigure{\thesection.\arabic{figure}}
\section{Additional Results}

\begin{figure}
    \centering
    \includegraphics[width=1.1\linewidth]{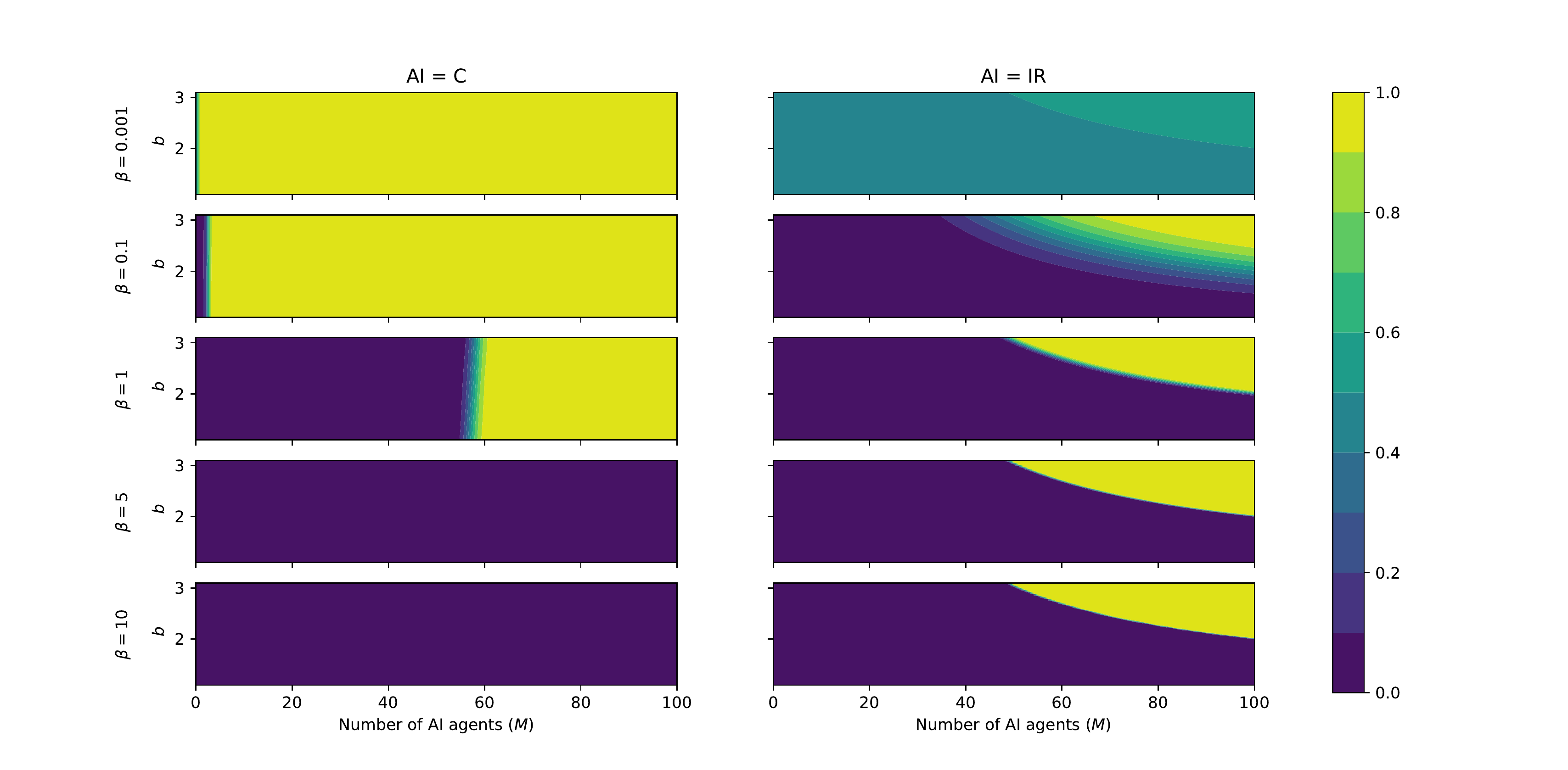}
    \caption{\textbf{Samaritan AI promotes higher levels of cooperation than discriminatory AI for weak intensities of selection, and vice versa for stronger intensities of selection.} Same as \autoref{anal_results_fin_pop} but for additional values of $\beta$.}
    \label{anal_results_fin_pop2}
\end{figure}

Figure \ref{anal_results_fin_pop2} shows additional results for other values of $\beta$.

\begin{figure}
    \centering
    \includegraphics[width = 1.1\linewidth]{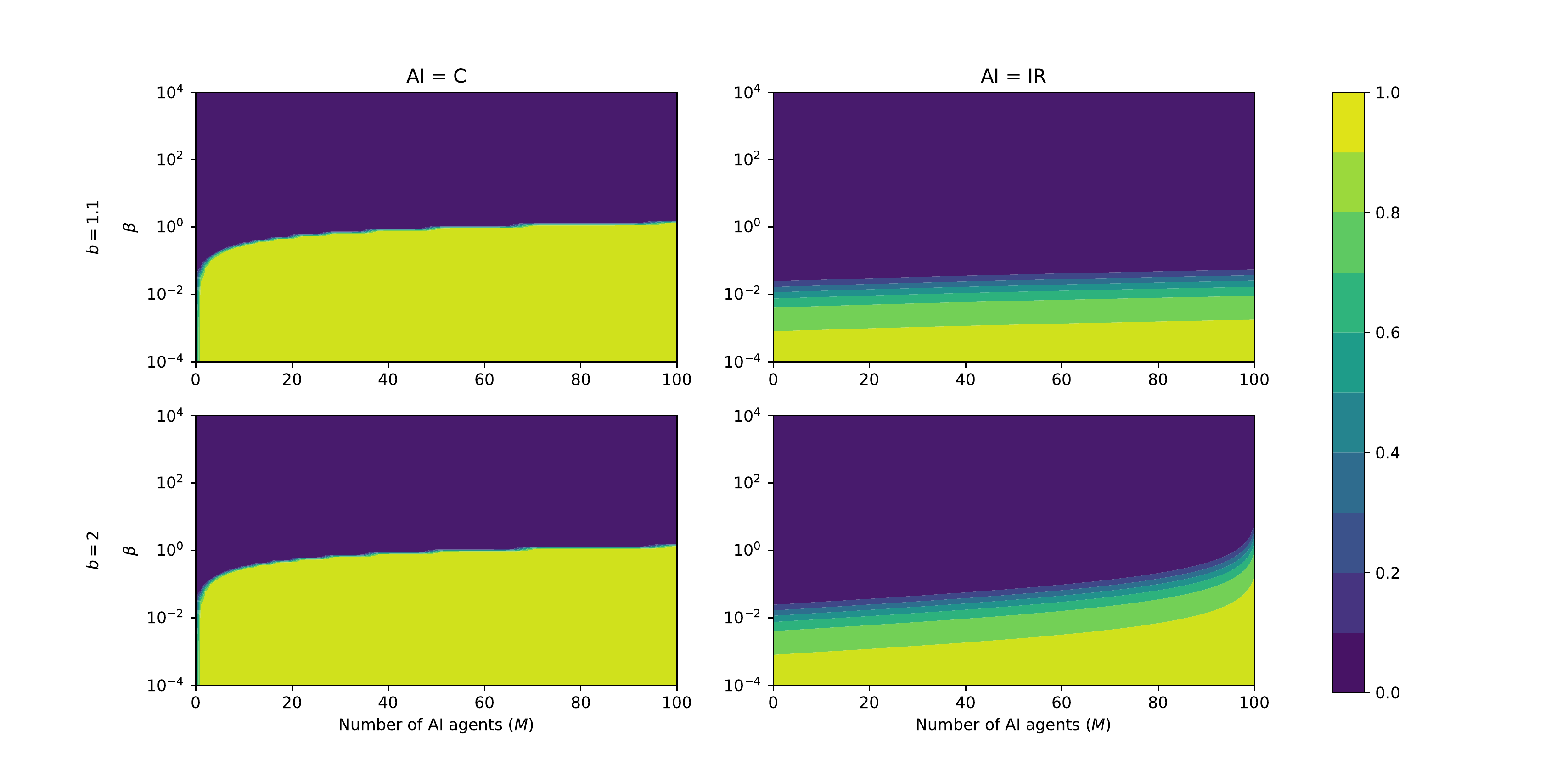}
    \caption{\textbf{Frequency of cooperation when fixing $b$ and varying $\beta$ and $M$.}}
    \label{anal_results_fin_pop3}
\end{figure}

Figure \ref{anal_results_fin_pop3} shows results for the frequency of cooperation for both types of AIs studied and for two fixed values of $b$.

\end{appendices}

%% file: sections/formulas.tex
\newpage
\section{Different tolerance for humans and AI}
We consider that if a human agent is aware that the role model is an AI agent, it might have a different probability of imitating the strategy  of AI. We model that by distinguishing between  human's and AI's  intensities of selection. They are now denoted as $\beta_H$ and $\beta_{AI}$, respectively. Thus, we rewrite 
\[
T^{+}(k) = \frac{N-k}{N}\left[\frac{k}{N+M}p_{-}(\beta_{H}) + \frac{\delta\cdot M}{N+M}p_{-}(\beta_{AI})\right]
\]

\[
T^{-}(k) = \frac{N-k}{N}\left[\frac{k}{N+M}p_{+}(\beta_{H}) + \frac{\delta\cdot M}{N+M}p_{+}(\beta_{AI})\right]
\]